\documentclass[aps]{revtex4}
\usepackage{graphicx,amsmath,color,bm}

\usepackage{setspace}   
\begin{document}

\title{I'm a doctor, not a mathematician! Homeostasis as a proportional--integral control system}

\author{Lennaert van Veen}
\email{lennaert.vanveen@uoit.ca}
\affiliation{Faculty of Science, Ontario Tech University, Oshawa, ON, Canada}

\author{Jacob Morra}
\email{jmorra@klick.com, Jacob.Morra@ontariotechu.net}
\affiliation{Labs Department, Klick Health, Klick Inc, Toronto, ON, Canada}

\author{Adam Palanica}
\email{apalanica@klick.com}
\affiliation{Labs Department, Klick Health, Klick Inc, Toronto, ON, Canada}

\author{Yan Fossat}
\email{yfossat@klick.com}
\affiliation{Labs Department, Klick Health, Klick Inc, Toronto, ON, Canada}

\begin{abstract}
  The distinction between ``healthy'' and ``unhealthy'' patients is commonly based on single,  discrete values taken at an isolated point in time (e.g., blood pressure or core temperature). Perhaps a more robust and insightful diagnosis can be obtained  by studying the functional interdependence of such indicators and the homeostasis that controls them. This requires quasi-continuous measurements and a procedure to map the data onto a parsimonious control model with a degree of universality. The current research illustrates this approach using glucose homeostasis as a target. Data were obtained from 41 healthy subjects wearing over-the-counter glucose monitors, and projected onto a simple proportional–integral (PI) controller, widely used in engineering applications. The indicators quantifying the control function are clustered for the great majority of subjects, while a few outliers exhibit less efficient homeostasis. Practical implications for healthcare and education are further discussed.
\end{abstract}

\maketitle

{\em Author Note: correspondence concerning this article should be addressed to Adam Palanica, Klick Labs, Klick Health, Klick Inc., 175 Bloor Street East, Suite 300, Toronto, Ontario, Canada, M4W 3R8. Phone: 416-214-4977. E-mail: apalanica@klick.com}

\section{Introduction}

How does one measure and assess health? {\em Measurement} may be referred to as quantifying attributes or characteristics belonging to a patient, whereas {\em assessment} may be referred to as drawing qualitative conclusions from the data that were measured \cite{CH}. Traditionally, the field of medicine has defined the traits that contribute to “health” as single, discrete values, or set ranges, often taken at a single time point \cite{B}. This is
true for many physiological functions, such as glycemia, core temperature, body mass index, bone density, cholesterol and blood pressure. These values are measured and assessed using simple scoring gradients where any patient whose value falls into a particular range may be defined as ``healthy'', and all others defined as ``unhealthy''.

Although using simple heuristics to measure and assess health may be efficient and unambiguous, this approach does not explain the fundamental control mechanisms and physiological systems that lead to
healthy values. Single values only measure the ``what'' of health and miss the ``how''. For example, a blood pressure of $120/80\, \text{mm Hg}$ may indicate a ``healthy'' value, but it is only taken at one time point in the patient's day. This value gives no indication of how effective the body is at controlling blood pressure when handling physical or mental stress.
In other words, the discrete, single time point values of physiological biometrics are merely manifestations of a deeper, more complex health control system.

Health may be better measured and assessed by studying the body's ability to maintain homeostasis, i.e., the maintenance of specific variables within an optimal range, regardless of external stimuli \cite{KM}. Many of today's most prevalent chronic illnesses, such as hypertension, diabetes, obesity, and depression, can be considered failures of the body's ability to maintain homeostasis or keep physiological signals within a normal working range. Therefore, knowing the functional model of healthy homeostasis may yield a better understanding of the overall well-being of the patient, and could become a fundamental step toward a more refined assessment of health and an early warning signal for illness.

To advance this idea, it is critical to examine a patient's individual homeostatic function and compare it to a base reference model to identify quantitative and qualitative deviations from healthy functioning.
However, two challenges currently render this approach impractical:
\begin{itemize}
\item existing models of homeostatic systems are complex;
\item the assessment of an individual's homeostasis function lacks a simple scoring system.
\end{itemize}
As a solution, we propose to describe the homeostatic function as a {\em control system} wholly independent of the underlying physiology. Control systems are widely used in engineering, economics, and cybernetics, and provide a strategy for maintaining the state of a system within a safe range without reference to its detailed mechanics. In particular, we consider a control strategy based on the system's current state and on its history. A scoring system can then be derived from the relative influence of these two factors for a given patient. 

One process that can be considered a control system is that of glucose homeostasis, whereas
a dysfunction of glucose homeostasis is associated with diabetes. It is estimated that more than 30 million Americans have diabetes, while another 84 million have prediabetes \cite{CDCP}. Diabetes is also associated with US\$327 billion of direct and indirect medical costs every year \cite{ADA}. Thus, an evaluation method to understand a patient's glycemic homeostatic function may be a key first step in reducing the economic and social burden of diabetes.

The standard methodology of measuring glycemic dysfunction includes HbA1C measurements, fasting blood glucose tests, and oral glucose tolerance tests. All three tests use simple heuristics to distinguish healthy patients from those with prediabetes or diabetes. Perhaps a better evaluation method to understand the nuanced structure of the glycemic system may be obtained by modelling its dynamic function. 
Although models of normal glycemic control currently exist, they tend to be fairly complicated. These models use a large number of variables and parameters, and describe a multitude of biophysical processes, rather than the resulting control strategy itself. For instance, the model recently proposed by Masroor {\sl et al.} \cite{MDABP} comprises 5 dynamical equations and over 25 parameters. The use of such models is limited by the {\em curse of dimensionality}, i.e., the catastrophic growth of the number combinations of parameter values to explore when attempting to reproduce measured data. We demonstrate that a simple model based on a feedback control widely used in engineering applications can robustly reproduce suitably pre-processed data. From the model parameters resulting from the fitting procedure, we extracted dimensionless indicators that quantify the homeostatic control function.


The procedure of fitting the model parameters relies on the availability of a quasi-continuous stream of data over a period much greater than the time scales typical for glucose production and consumption. Recently, 
off-the-shelf continuous glucose monitoring (CGM) technology has become available to provide a convenient and cost-effective way to accurately measure continuous glycemia from subjects pursuing their regular daily activities. The present study utilized the FreeStyle Libre glucose monitor (Abbott Diabetes Care) on participants not diagnosed with any medical condition to gather glucose level data
every 15 minutes for 2 weeks. From these data, we extracted a sequence of glucose levels representative of the subject's feedback control, and then iteratively optimized the model parameters to reproduce it. Once the model is developed, this fitting procedure would only take a few seconds on a single modern central processing unit to execute.

The objectives of the current research were threefold:
\begin{itemize}
\item to extract a personalized, functional description of a subject's glucose homeostasis from easily obtained, quasi-continuous measurements;
\item to confirm the universality across subjects of the homeostatic feedback system model;
\item to extract medically actionable indicators from this functional description.
\end{itemize}

\section{\label{sec:data}Methods and materials}

\subsection{\label{subsec:participants}Participants}

Data were collected from 41 participants (20 females; 21 males; age range = 19--50 years, M age = 32.4 years, SD = 6.8 years). Participant race included 23 (56.1\%) Caucasian, 15 (36.6\%) Asian, 1 (2.4\%) African American, 1 (2.4\%) Hispanic, and 1 (2.4\%) mixed race (Caucasian and African American).
Exclusion criteria were participants below the age of 18, those who were diagnosed with any mental or physical medical condition of any kind (chronic or acute), those taking any form of prescription medication, and those who were pregnant or breastfeeding. This sample of participants had an average body mass index of 25.8 (SD = 5.7), an average resting blood pressure of 120/75 mm Hg, and an average resting heart rate of 72 bpm.

The study took place at Klick Inc., which is a technology, media, and research company in the healthcare sector based in Toronto, Canada. All of the participants were employees of Klick Inc. The study was performed in accordance with relevant guidelines and regulations, and all participants signed informed consent. The study received full ethics approval from Advarra IRB Services (www.advarra.com/services/irb-services/).

\subsection{\label{subsec:apparatus}Apparatus}

The FreeStyle Libre flash glucose monitoring system (Abbott Diabetes Care) was used to measure real-time, continuous interstitial glucose levels with a minimally invasive 5 mm flexible filament inserted into the posterior upper arm. The sensor works based on the glucose-oxidase process by measuring an electrical current proportional to the concentration of glucose. The FreeStyle Libre is a factory calibrated device, designed not to require finger prick tests during use. Previous research has shown the FreeStyle Libre to have consistent accuracy and reliability throughout the 14 days with a mean absolute relative difference of $11.4\%$ compared with capillary blood glucose, and is not significantly influenced by age, sex, body weight, BMI, or time of use (day versus night) \cite{HBLC,BBCKA,EACHMRRST}.

The device contains a sensor which is attached to the posterior region of the upper arm with an adhesive patch, and a handheld reader device which downloads data from the sensor via near-field communication. Interstitial glucose concentrations (in mmol/L) are captured by the sensor every 15 min and/or when users scan the sensor using the handheld device. The handheld device requires users to scan the sensor at least every 8 hours, otherwise previous data are overwritten by the sensor. The system has a lifespan that restricts sensor wear to 14 consecutive days, after which the handheld device will no longer download data from the sensor. In our particular sample, the glucose sensors lasted an average of 13.0 days (due to some cases of malfunction or detachment), with a range of 7 to 14 days.

\subsection{\label{subsec:collection}Data collection}

At the beginning of the study period, participants completed a self-report demographic survey, and had some physiological variables measured, including height, weight, body mass index (BMI), resting blood pressure, and resting heart rate. Participants were then outfitted with the FreeStyle Libre flash glucose monitor, and instructed on its use. Participants were instructed to scan the sensor with the handheld device at least once every 8 hours to minimize data loss. Missing data were anticipated as participants may have slept over 8 hours, so they were encouraged to scan the device before going to sleep and immediately upon waking. Other than using the glucose device, no other intervention was implemented, and participants were not asked to change their lifestyle in any way.

\subsection{\label{subsec:availability}Data availability}

The data that support the findings of this study are available from the corresponding
author upon reasonable request.

\section{\label{sec:model}A simple control model}

Feedback control is a strategy to minimize the deviation of a process variable, in our case the glucose level, from a set value. A simple strategy, refered to as proportional–integral (PI) control, is
based on a response proportional to this deviation and on an integral over its history. The application of PI control goes back at least as far as 1922, when
N. Minorsky proposed to use it for the automatic steering of ships. Since then, it has been used to control processes as diverse as the pasteurization of milk at a
constant temperature, and the balancing of flying drones \cite{B2,AAMR}. We conjecture that the PI controller can effectively describe the homeostatic control system 
resulting from various physiological pathways, independently from the details of any of these pathways. Our assumptions are the following:
\begin{enumerate}
\item There is an instantaneous response in proportion to the deviation of the glucose level from the set point, for instance through the release of insulin or glucagon into the blood stream.
\item There is memory in the system due to the finite time it takes the body to excrete or metabolize the hormones involved. It is reasonable to assume that the rate at which this happens is proportional to the hormone concentration, so that the memory fades exponentially in time.
\end{enumerate}
The PI controller is coupled to a rudimentary model of blood glucose kinetics. It comprises only the effects of the base metabolic rate, food intake, and the control feedback. The feedback term takes the form of {\em mass action kinetics}. That is, it corresponds to the rate of a hypothetical chemical reaction between two substances with concentrations $u$, the control rate, and $(e+e_{\rm sp})$, the total glucose concentration, in a well-mixed reaction vessel under the assumption that the reaction takes place with constant probability every time different molecules collide. This form of the control term coincides with that proposed by Bergman {\sl et al.} \cite{BIBC}, who tested several models of insulin-glucose interactions. In that study, a measured quantity of insulin was injected directly into the blood stream, and the insulin and glucose concentrations were measured at regular intervals after that. The interaction term proportional to both the insulin and the glucose concentration was shown to best model the study.

If $e$ denotes the deviation from the set point glucose concentration, $e_{\rm sp}$, the equations are
\begin{align}
  u &= A_1 e + A_2 \int\limits_{t'=-\infty}^{\infty} w(t-t',\lambda) e(t')\,\mbox{d}t', \label{PI}\\
  \frac{\mbox{d} e}{\mbox{d}t} &= -A_3 +F(t) - u\, (e+e_{\rm sp}),  \label{dedt}\\
  w(\tau,\lambda) &=\begin{cases} 0 &\text{if}\ \tau<0 \\ \lambda \exp(-\lambda \tau) &\text{if}\ \tau > 0 \end{cases}. \label{weight}
\end{align}
Here, $F(t)$ models the release of glucose into the blood stream and
$w$ models the sensitivity of the control variable to past glucose concentrations. We measure $e$ and $e_{\rm sp}$ in units mmol/litre, while $u$ has
units $1/\Delta$, where $\Delta=15\,\text{mins}$ is the interval of measurements taken by the monitoring system described in the Apparatus section.
The parameters of the model are summarized in Table 1.

\begin{table}
  \begin{tabular}{|l|l|l|l|}\hline
    parameter & meaning & range & units \\\hline
    $A_1$ & Proportional control term & $-0.1$ -- $0.7$ & $\text{litre}/(\Delta \times \text{mmol})$ \\\hline
    $A_2$ & Integral control term & $0.09$ -- $0.75$ & $\text{litre}/(\Delta \times \text{mmol})$ \\\hline
    $A_3$ & basic metabolic rate & $0.005$ & $\text{mmol}/(\Delta \times \text{litre})$ \\\hline
    $\lambda$ & decay rate of the Integral term & $0.1$ -- $0.7$ & $1/\Delta$ \\\hline
  \end{tabular}
  \caption{\label{parameters}Parameters of the glucose homeostasis model with their meaning and typical range across test subjects. Here, $\Delta=15\, \text{minutes}$ is the interval between two measurements of the FreeStyle Libre flash glucose monitor.}
\end{table}

For the equation, we assigned $A_3=0.005\,\text{mmol}/(\Delta \times \text{litre})$, where $\Delta=15\,\text{(mins)}$ and the time in between automated measurement from the FreeStyle Libre
device. This corresponds to a base metabolic rate of about $1300$--$1950$ (Kcal) per 24 hours and a blood volume of about $4.5$--$6.4$ (litres).

\section{\label{sec:fitting}Reproducing measured data with the model}

For each subject, the parameter values of the PI model's equations (\ref{PI}--\ref{weight}) are chosen to minimize the difference between the model output and measured glucose data. We did not fit the measured raw data as it may be variable due to noise. Instead, we selected a number of peaks of glucose and used their average as the representative peak for the subject. Averaging over too few data segments yields a representative peak with too much noise, while averaging over too many obfuscates the structure of the data. We found that three to five sample peaks were sufficient. Figure 1 shows 
\begin{figure}[t]
\includegraphics[width=0.495\textwidth]{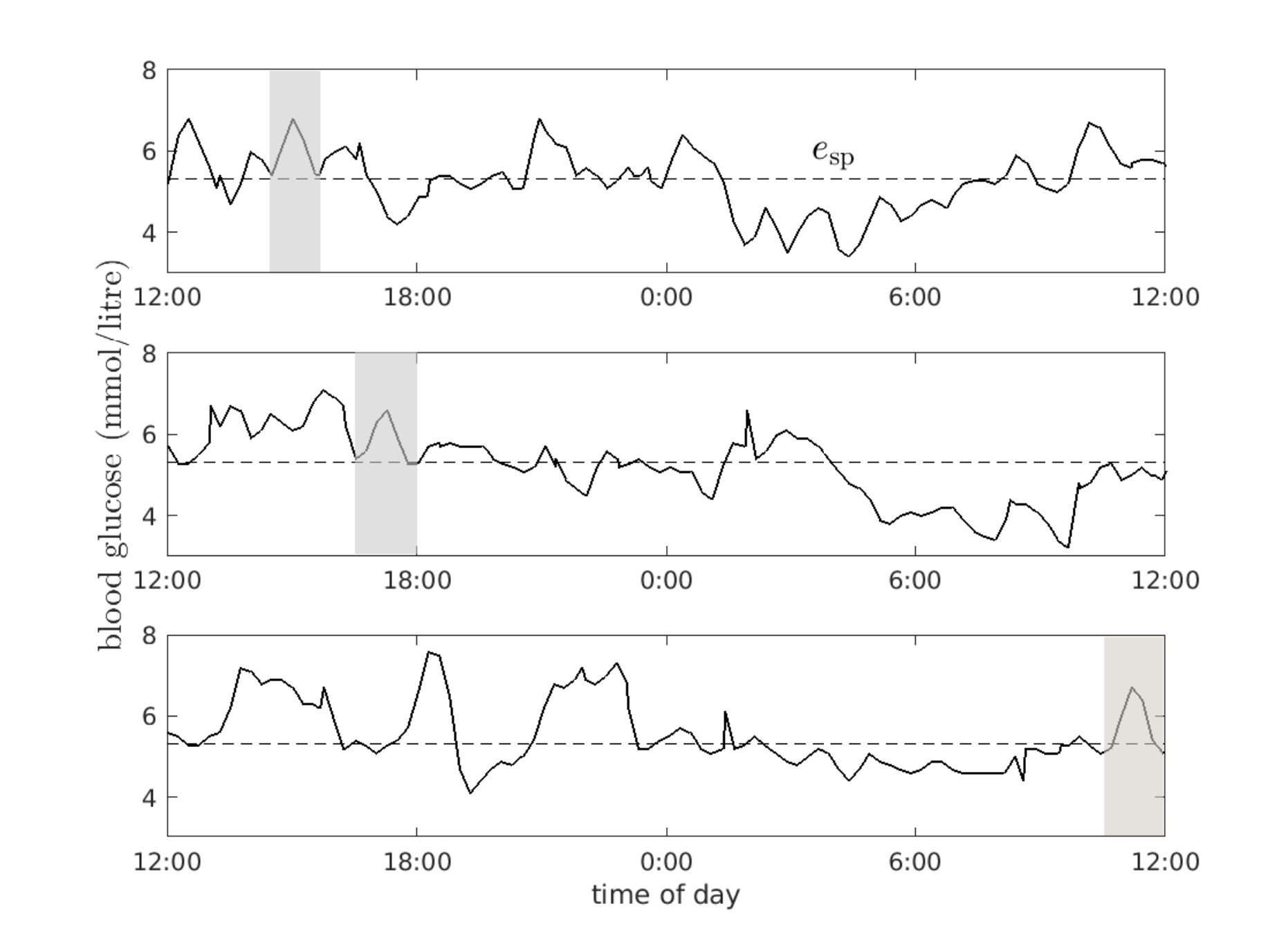} \includegraphics[width=0.495\textwidth]{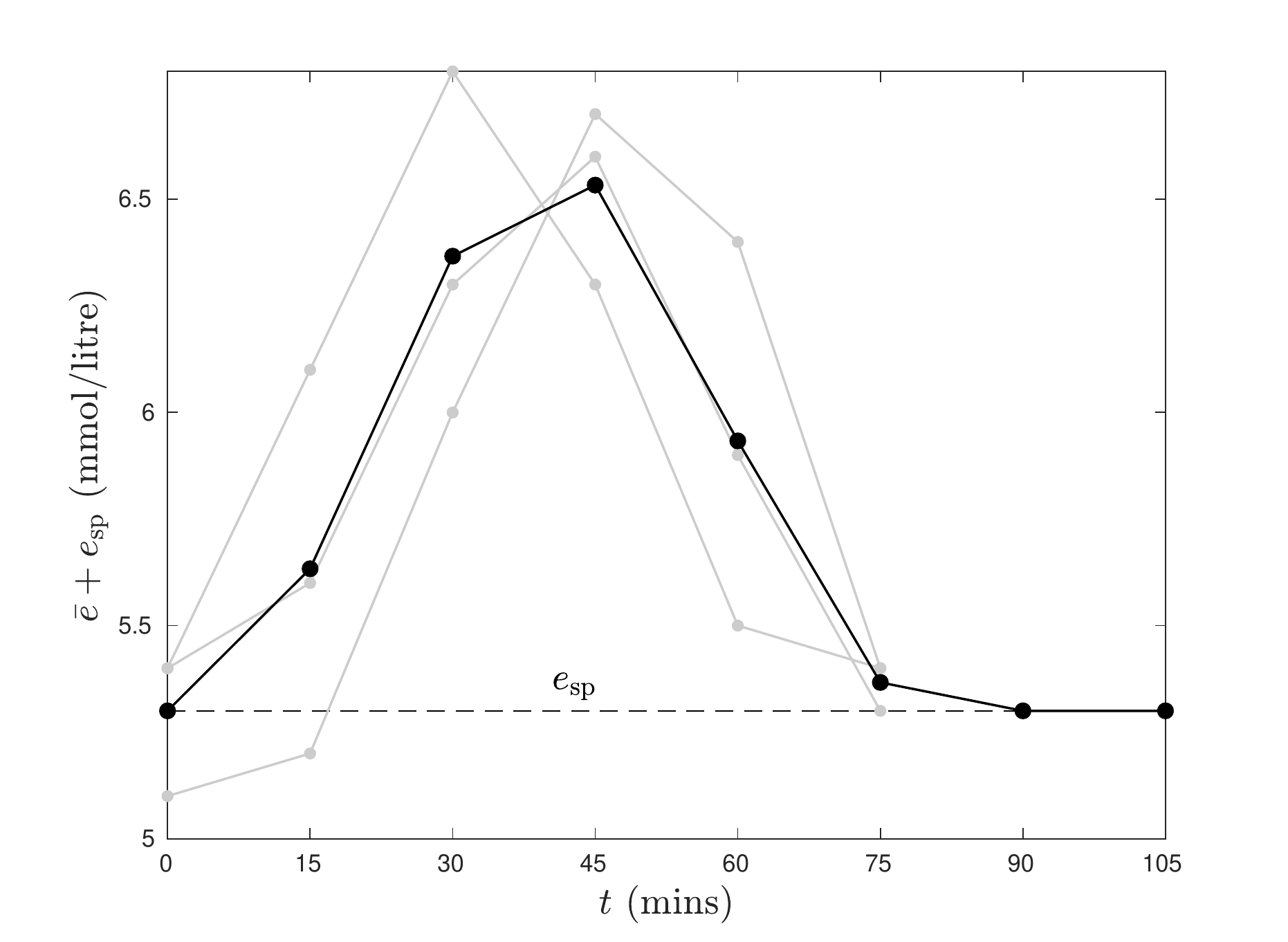}
\caption{\label{rep_peak}Construction of the representative peak for one subject. Left: three days of raw data. The shaded time segments were averaged to find a representative peak. The dashed line corresponds to the set point glucose level $e_{\rm sp}$. Right: the three selected peaks (grey) and the representative peak (black). The set point is taken to be the minimum over the representative peak. Two data points at the set point are added to the representative peak at the end. The fit for this representative peak is shown in the rightmost inlay in Figure \ref{scatter}.}
\end{figure} raw data for one test subject. Three peaks in glucose, indicated by shading, have been selected to form the representative peak, $\bar{e}$, shown on the right of Figure 1; the set point glucose level, $e_{\rm sp}$, corresponds to the minimum over the representative peak and is indicated by a dashed line.

Once the representative peak has been distilled from the raw data of a given subject, we tune the model parameters to reproduce it as accurately as possible. The accuracy of a fit is measured by the function
\begin{equation}\label{cost_function}
E= \frac{\sum_{i=1}^{n_{\rm peak}} (\bar{e}(t_i)-e(t_i))^2}{\sum_{i=1}^{n_{\rm peak}} \bar{e}(t_i)^2}
\end{equation}
where $n_{\rm peak}$ is the number of points in the representative peak, $\Delta=15$ (mins) apart. This least-squares fit is computed by a steepest descent algorithm, the details of which are explained in the supplementary material.

\section{\label{sec:results}Results}

We computed the best fit for each of the 41 subjects, reaching a residual mismatch of $E<0.06$ for all; for $90\%$ of the subjects, the mismatch was less than $0.02$. A scatter plot of the results are shown in Figure 2. On the axes are the dimensionless indicators $\sigma_e A_1/u_{\rm m}$ and $\sigma_e A_2/u_{\rm m}$, where $\sigma_e$ is the standard deviation of all glucose measurements for a given subject and $u_{\rm m}$ is the maximum attained by the control variable in the optimal fit. The data for most participants are clustered in the range $0.3 \pm 0.25$ ($78\%$) for both indicators, and there is a notable group of outliers with $\sigma_e A_1/u_{\rm m}<0$ and $\sigma_e A_2/u_{\rm m}>0.5$. Three inlays have been included to illustrate the qualitative difference between the fitted curves. Inlay A shows the data point close to the mean value for both indicators ($0.22$ and $0.40$, respectively). The width of the model output peak $e(t)$ was about $50$ (mins) and the control variable has a smooth peak, delayed by about $15$ (mins), and decays back to zero about $1$ hour after the glucose peak. If we take the control variable $u$ to be a proxy for the insulin concentration, these numbers agree well with data for healthy subjects undergoing a meal glucose tolerance test presented by Caumo {\sl et al.} \cite{CBC}. 

Inlay B shows a more efficient response to approximately the same input. While the amplitude of the input function is close to that in inlay A, the peak glucose level is almost two times lower. The width of this peak is only about 30 (mins) and the control variable assumes its resting value within 30 (mins) from the glucose peak value.

On the other end of the scale, inlay C shows a relatively inefficient response. The measured data exhibit a plateau at a level of over 2 (mmol/litre) in excess of the set point glucose level. While the model accurately captures the rapid rise over the first 30 (mins), the measured and modelled data diverge somewhat over this plateau. The control variable reaches a level that is about four times higher than that in inlays A and B and remains high after the measured and modelled glucose concentrations return to the set point.

These qualitative differences can be understood from the structure of the model. If $A_1,\,A_2 >0$ and $A_1 > A_2$, the controller (\ref{PI}) is mostly determined by an instantaneous, proportional response. For a rapidly fluctuating glucose level, this may lead to fluctuations of the control feedback that cannot be sustained by any physiological mechanism, but for a peak generated by a regular meal, it leads to a quick reset. On the other side of the diagram, if $A_1<0$ and $A_2>0$ the integral and proportional terms of the control model have an opposite effect. This is the only way for our model to produce a sustained excess glucose level. In the first phase, when the representative peak is rapidly rising, the proportional term is dominant and gives rise to a positive feedback. This pushes the model glucose level up rapidly. The duration of this phase is about $1/\lambda$. In the second phase, the representative peak reaches a plateau and the balance between the proportional and integral terms is set by the difference $A_1-A_2$. This difference must be negative in order for the glucose level to decrease, if only rather slowly.
\begin{figure}
\includegraphics[width=\textwidth]{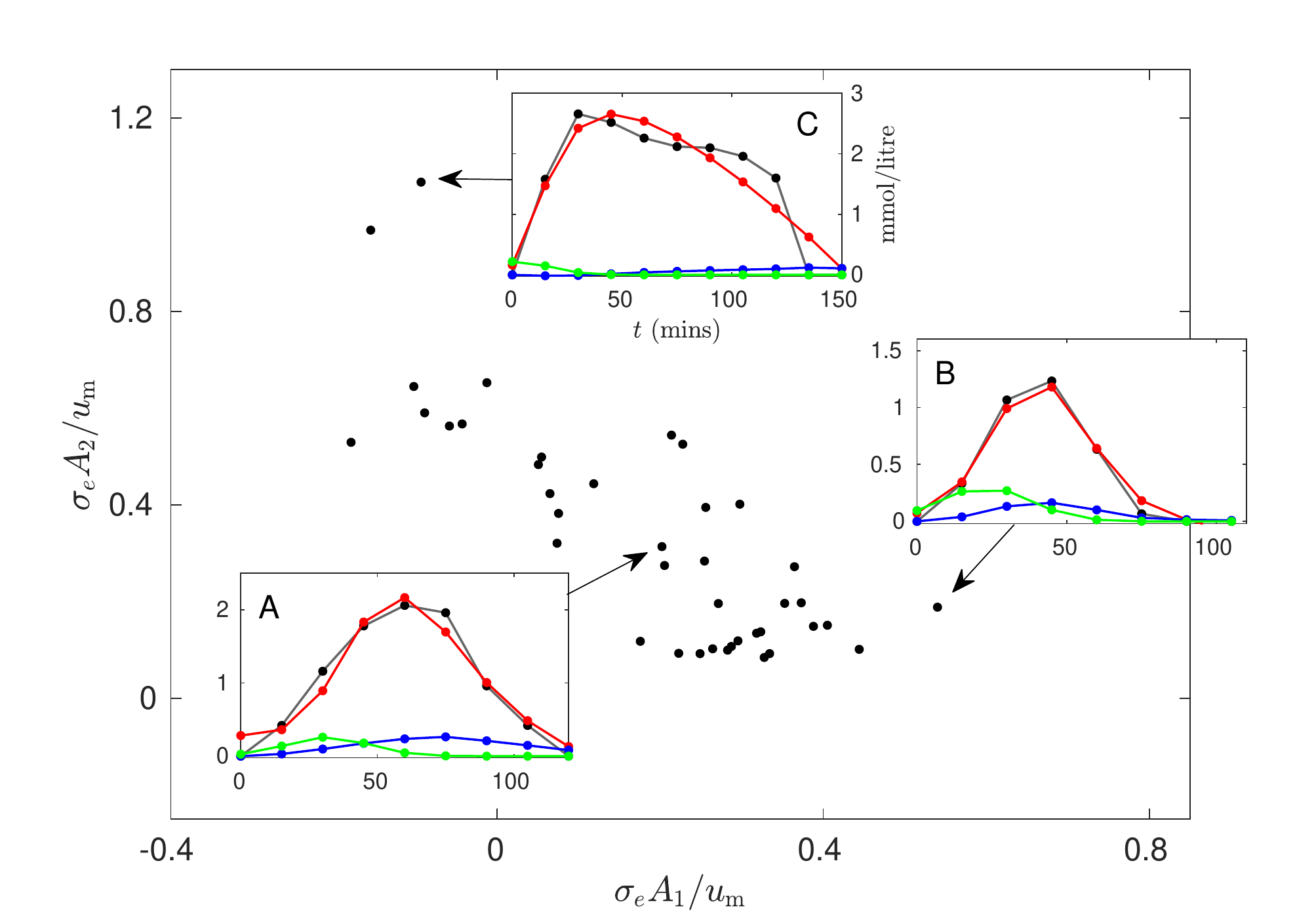}
\caption{\label{scatter}Scatter plot of the optimal model parameters for all subjects. Shown are $A_1$ and $A_2$, nondimensionalized by the standard deviation of each subject's time series of the blood glucose level, $\sigma_e$, and the corresponding maximum of the control variable, $u_{\rm m}$. Three illustrative results of the fitting procedure are shown in inlays. In these, the black lines correspond to the representative peaks and the red lines to the output $e(t)$ of the control model (\ref{PI})--(\ref{weight}). The normalized input function, $F(t)/\lambda$, and the normalized control variable, $\sigma_{e}u(t)/\lambda$, are shown in green and blue, respectively. }
\end{figure}

\section{\label{sec:discussion}Discussion}

The goal of this research was to propose a new technique of health measurement and assessment by examining homeostatic control systems rather than discrete, single values. We also wanted to improve existing models of homeostatic systems by developing uncomplicated algorithms for easily extracting medically relevant insights and actions. In this case, our aim was to formulate a model of glucose homeostasis that emphasized simplicity and universality versus physiological detail. {\em Simple} refers to the fact that our model has only two variables (the control variable $u$ and the excess glucose concentration $e$) and four parameters (those that regulate the control strategy, $A_1$ and $A_2$, as well as the amplitude and shift of the input function). {\em Universal} refers to the fact that the model can be applied to all test subjects under differing circumstances, while no explicit reference is made to the pathways of glucose control, or potentially to other forms of homeostasis, like blood pressure or core temperature.

While a significant body of research exists in modelling glucose homeostasis \cite{PDBG}, the mathematical models that currently exist for humans or animals include a multitude of variables and physiological parameters. In the case of machine learning-based models, the result is individualised to each subject \cite{JFBLFW}. These features limit their translation to clinical settings where simplicity, interpretability, and universality are paramount.

The current research demonstrates that the technique for processing glucose data from a portable device has low computational complexity, and can, in principle, be done in real-time. For example, this process would consist of a few straightforward steps:
\begin{enumerate}
\item From the raw data, 3 -- 5 peaks are selected of comparable width. The average of these is the representative peak for the subject in question.
\item The model parameters are iteratively tuned to make the model output as similar as possible to the representative peak.
\item From the optimized parameters and the output control variable, two dimensionless indicators are extracted that encode the efficiency of the control system.
\end{enumerate}

The PI control strategy can provide a sophisticated model for examining glucose homeostasis in humans. The indicators that are extracted give more detailed information than single, discrete values, like HbA1C, since they present information about the way a subject's glucose level is being controlled rather than with statistics on the glucose level itself. The mathematical techniques are also similar in other fields, such as biophysics, where {\em synchronization} can be found in a variety of systems, from firing neurons \cite{MGN} to signalling fireflies \cite{CM}. Similarly, pattern formation near critical transitions occurs in the same way across a range of applications, from electrical signals in cardiac tissue to density variations in bacterial populations. Thus, models based on PI control could yield a parsimonious description of a variety of homeostatic functions, regardless of their particular mechanics. 

We demonstrated that the extracted indicators fall within a well-defined range for $78\%$ of the subjects, while there are a few outliers for which the representative peak shows an unusually slow decay.
For these outliers, the proportional and integral terms of the control strategy work against each other.
We may speculate whether
this indicates a pathological state
such as prediabetes, but future research is needed to investigate this possibility; the current research utilized participants who were not diagnosed with any medical condition. Nevertheless, if it is the case that this PI control model does predict or diagnose a some type of medical issue, it would give extra credibility to its usefulness in everyday practice.

Since this pilot study was exploratory in nature, there are some limitations which should be addressed with further research. Firstly, since the current data was collected from individuals not diagnosed with any medical condition, the indicators that rate the effectiveness of the controller may need to be refined in a sample of patients diagnosed with prediabetes or type 2 diabetes for comparison. Secondly, we have assumed that the input peak $F(t)$ always takes a Gaussian form, since this assumption agrees reasonably well with data measured {\em in vitro} \cite{MMV}. However, in order to turn our data processing pipeline into a diagnostic tool, we may need to allow for a wider class of input functions; more a priori knowledge of the food intake, for instance, would also help to model the release of glucose into the blood stream over time. Lastly, this study used the FreeStyle Libre glucose monitor due to convenience, as this study was performed in Canada, where the device is available without any prescription; the FreeStyle Libre measures glucose in the interstitial fluid (ISF), but not directly in the blood. While the glucose levels in the ISF are closely inline with blood glucose, a delay of 5-10 minutes has been estimated by various studies \cite{RSS}. Nevertheless, this study shows that an off-the-shelf consumer level device may be used to redefine homeostasis measurement and assessment.

This study contributes to the field of digital medicine by providing an improved method of understanding health beyond single values, and furthering our understanding of homeostasis in the normal population. Importantly, these techniques can be utilized in other patient populations and potential disease states. These findings have practical implications for healthcare and education where enhanced translation of medical knowledge can empower both the physician and patient.

\section*{Acknowledgments}

This study was partially funded by NSERC–Engage as a collaboration between L.v.V and Y.F. We would like to thank all of the participants, project members, supporters, and researchers at Ontario Tech University and at Klick Inc. for the successful development, implementation, and evaluation of this research. This research was internally funded and received no specific grant from any funding agency in the public, commercial, or not-for-profit sectors.

\section*{Competing Interests}

The authors declare that they have no conflict of interest.

\section*{Author contributions}

L.v.V. supervised the data analysis, developed the mathematical model, and contributed to writing the manuscript. J.M. conducted the data analysis, and helped interpret the results. A.P. executed the study, collected the data, helped with data analysis, and contributed to writing the manuscript. Y.F. designed the study, assisted in data analysis, and contributed to the writing and editing of the manuscript. All authors contributed to the final review and editing, and have approved the final manuscript.

\newpage
\centerline{\rule{0.2\textwidth}{2pt}\ \ {\bf Supplementary material}\ \ \rule{0.2\textwidth}{2pt}}

\appendix
\section{\label{Fit} Details of the parameter fitting}

The parameters $A_1$, $A_2$ and $\lambda$ in control model (\ref{PI}), as well as the amplitude and timing of the input peak $F(t)$ in the bood glucose model (\ref{dedt}), are
  specific to each representative peak, extracted from a subject's sequence of measurements. We find the least-squares fit by a simple gradient descent algorithm that
  relies on the simulation of $u(t)$ and $e(t)$ over a time interval as long as the representative peak.
\begin{figure}[t]
\includegraphics[width=\textwidth]{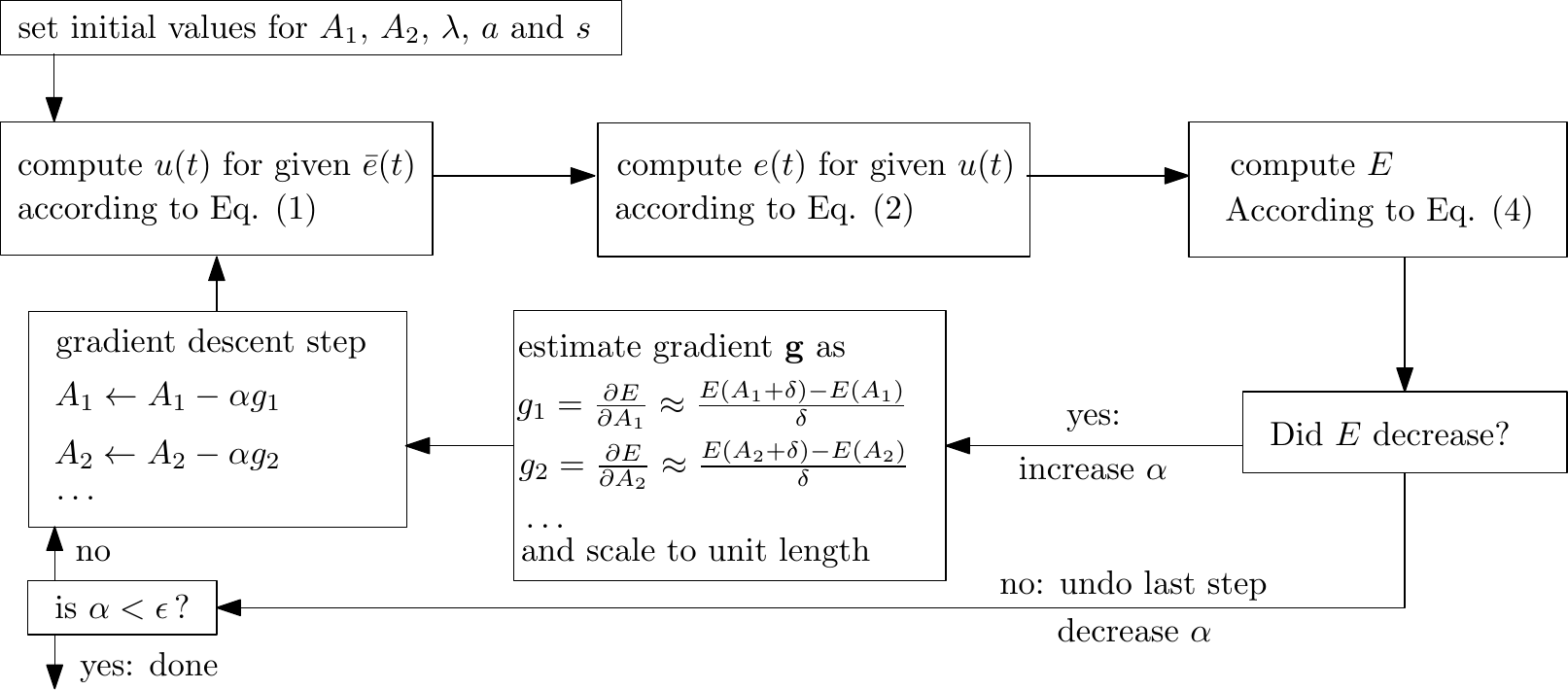}
\caption{\label{schematic}Schematic representation of the parameter fitting procedure. Starting at the top left, we set initial values for the tunable parameters. We then compute the model output $e(t)$ and its difference from the representative peak $\bar{e}$. The gradient of the function $E$, which measures the mismatch, is approximated by finite differencing with a small constant $\delta$ making a $1\%$ variation of the parameter. A gradient descent step of size $\alpha$ is then taken. If $E$ increases from one iteration to the next, the step is rejected and $\alpha$ is decreased. When $\alpha$ is smaller than a pre-set threshold $\epsilon<10^{-12}$, the algorithm has converged.}
\end{figure}

The gradient descent algorithm is illustrated in Fig. \ref{schematic}. For given values of $A_1$, $A_2$ and $\lambda$, we compute a time series of the control variable $u(t)$ from $\bar{e}(t)$ according to Eq. (\ref{PI}). Next, we obtain the model output $e(t)$ from the control variable and the input function $F(t)$ according to Eq. (\ref{dedt}). We assume the input function to take the form of a Gaussian peak with amplitude $a$, standard deviation $\Delta$ and the maximum at $t=s$. While the shape of the input peak does not strongly influence the fitting procedure, its amplitude and the shift of its peak with respect to that of the representative data $\bar{e}$ do. Since we have no a-priori estimates for these quantities, we add them to the list of parameters to be optimized. In addition, the initial value $e(0)$ can be added to this list, but the results remain qualitatively the same if we fix $e(0)=\bar{e}(0)$ for each subject.

From the representative data and the model data, we then compute $E$ according to Eq. (\ref{cost_function}). Next, we approximate the gradient $\bm{g}$ of $E$ with respect to the tunable parameters $A_1$, $A_2$, $\lambda$, $a$ and $s$ by a simple finite difference approximation and scale it to unit length. We then take a gradient descent step in the direction $-\mathbf{g}$ of size $\alpha$. The step size $\alpha$ is reduced if $E$ does not decrease from one iteration to the next, and increased if it does. 
The optimal parameter values are found when $\alpha$ decreases below some pre-set threshold.

\end{document}